\documentclass[11pt]{article}
\usepackage{moriond,epsfig}

\def\Journal#1#2#3#4{{#1} {\bf #2}, #3 (#4)}

\def\NPA{{\em Nucl. Phys.} A}
\def\NPB{{\em Nucl. Phys.} B}
\def\PLB{{\em Phys. Lett.}  B}
\def\PRL{\em Phys. Rev. Lett.}
\def\PRD{{\em Phys. Rev.} D}

\def\be{\begin{equation}}
\def\ee{\end{equation}}
\def\bea{\begin{eqnarray}}
\def\eea{\end{eqnarray}}

\begin{document}
\vspace*{4cm}
\title{GLUON RADIATION OFF MASSIVE QUARKS IN A QCD MEDIUM}

\author{\underline{N\'estor Armesto}, Carlos A. Salgado and Urs Achim
Wiedemann}

\address{Department of Physics, Theory Division, CERN,
CH-1211 Gen\`eve 23, Switzerland}

\maketitle\abstracts{Medium-induced gluon radiation from massless and massive
quarks is treated in the same formalism. The dead cone which regulates gluon
radiation from massive quarks in the vacuum at small angles,
is filled in the medium but
constitutes a small fraction of the available phase space. Our study indicates
that the energy loss for charmed hadrons at RHIC should be smaller
than for light hadrons, but still sizable.}

\section{Introduction}
Gluon radiation is the dominant process for energy loss of high-energy
partons traversing a
strongly interacting medium (see~\cite{reviews} for some reviews).
It implies the energy degradation of the leading parton, the broadening of its
associated
parton shower and the increase of the associated hadron multiplicity.
Evidences for this mechanism for light partons
have been obtained in Au+Au collisions at
RHIC (see~\cite{recent} and references therein).
The question we address here is how the
medium-induced gluon
radiation off a massive quark
differs from that off a massless parton; full details can be found
in~\cite{noso}.

The conventional formalism describes the medium-modification of the
vacuum radiation pattern taking into account all possible rescatterings of
the incoming and outgoing partons~\cite{reviews}.
In the absence of a medium it reproduces the
known results for radiation in the vacuum: for massless quarks it leads to
$\omega\frac{dI_{\rm vacuum}}{d\omega\, d{\bf k}_\perp}\propto {1 \over
{\bf k}_\perp^2}$, with $\omega$ the energy and ${\bf k}_\perp$ the
transverse momentum of the emitted gluon.
On the other hand, it is well known
that gluon radiation in the vacuum is modified by
a mass of the parent quark: radiation for angles $\theta<m/E$ is
suppressed,
the so-called dead cone effect~\cite{dce}.
It turns out that the ${\bf k}_\perp^{-2}$ singularity is changed into
${1\over
{\bf k}_\perp^2} \left[{{\bf k}_\perp^2 \over
{\bf k}_\perp^2+\left({m\omega\over E}\right)^2}\right]^2
\equiv {1\over
{\bf k}_\perp^2}\ F\left({\bf k}^2_\perp,{m\omega \over E}\right)$, with
$F\left({\bf k}^2_\perp,{m\omega \over E}\right)$ the dead cone factor.
In a first exploratory study, Dokshitzer and Kharzeev~\cite{dk} proposed that
medium-induced gluon radiation
is reduced by the same effect. They considered
\begin{equation}
\label{eq1}
\omega\frac{dI^{m>0}_{\rm medium}}{d\omega} =
\omega\frac{dI^{m=0}_{\rm medium}}{d\omega}\ F\left(\langle{\bf
k}^2_\perp\rangle,{m\omega \over E}\right),
\end{equation}
with $\langle{\bf
k}^2_\perp\rangle$ some average gluon transverse momentum squared.

\begin{figure}
\begin{center}
\psfig{figure=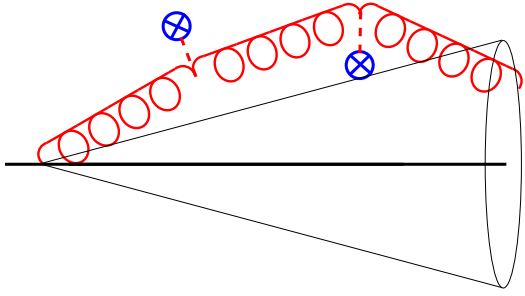,height=1.5in}
\end{center}
\caption{Diagram showing the filling of the dead cone due to rescattering of
the radiated gluon.
\label{fig1}}
\end{figure}

Working in the formalism of~\cite{wiedemann}, we get~\cite{noso}
\begin{eqnarray}
& \omega\frac{dI}{d\omega\, d{\bf k}_\perp}=\omega\frac{dI_{\rm medium}}{d\omega\, d{\bf
k}_\perp}+\omega\frac{dI_{\rm vacuum}}{d\omega\, d{\bf k}_\perp}
  = {\alpha_s\,  C_F\over (2\pi)^2\, \omega^2}\,
    2{\rm Re} \int_{0}^{\infty}dy_l
  \int_{y_l}^{\infty}  d\bar{y}_l\,
  e^{i \bar{q} (y_l - \bar{y}_l)}
   \int d{\bf u}\, e^{-i{\bf k}_\perp \cdot {\bf u}}&
    \nonumber \\
&\times\ \
  e^{ -\frac{1}{2} \int_{\bar{y}_l}^{\infty} d\xi\, n(\xi)\,
    \sigma({\bf u})} \,
   {\partial \over \partial {\bf y}}\cdot
  {\partial \over \partial {\bf u}}\,
  \int_{{\bf y}=0={\bf r}(y_l)}^{{\bf u}={\bf r}(\bar{y}_l)}
  {\cal D}{\bf r}
   \exp\left[ i \int_{y_l}^{\bar y_l} \hspace{-0.2cm} d\xi
        \frac{\omega}{2} \left(\dot{\bf r}^2
          - \frac{n(\xi) \sigma\left({\bf r}\right)}{i\, \omega} \right
)\right].&
\label{eq2}
\end{eqnarray}
Here the information about
the medium is contained in the product density times cross section, for which
we take the multiple soft scattering approximation
$n(\xi)\sigma\left({\bf r}\right)\simeq {1\over 2}\  \hat{q}(\xi) {\bf
r}^2$ (see~\cite{reviews,noso,wiedemann}).
The only model parameter is $\hat{q}(\xi)\simeq \langle {\bf q}^2_\perp
\rangle_{\rm medium} / \lambda$, i.e. the average transverse
momentum squared transferred from the medium per mean free path length.
We take $C_F\alpha_s=4/9$.
We have checked, paralleling for the
massive case
the derivation in~\cite{wiedemann},
that all mass effects go into the exponential
containing the difference of three-momenta of the incoming and outgoing
partons $\bar{q} = p_1 - p_2 - k \simeq
\frac{x^2\, m^2}{2 \omega}$, $x=\omega/E\ll 1$, with $E$ the energy of the
radiating parton.
In this formalism the dead cone for the vacuum is recovered.
On the other hand one would expect naively
that for
the medium term, the dead cone effect
will be absent, see Fig.~\ref{fig1}. Technically
this will
result from a competition between interference and rescattering.

\section{Results}
\begin{figure}
\begin{center}
\psfig{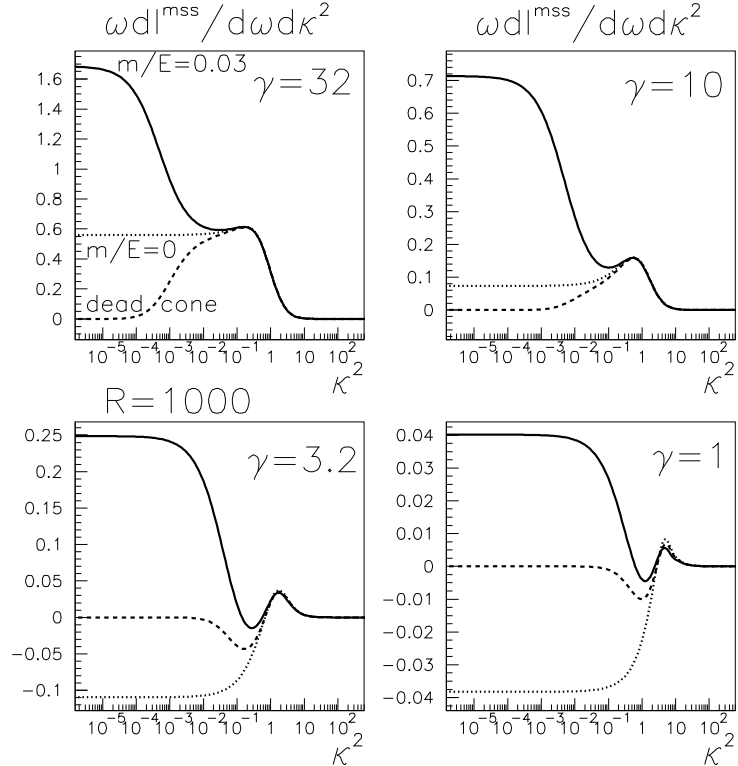}
\end{center}
\caption{Gluon energy distribution versus $\kappa^2={\bf
k}^2_\perp/(\hat{q}L)$ for $m/E=0.03$ and
different gluon energies $\gamma=\omega_c/\omega$.
\label{fig2}}
\end{figure}

It turns out to be convenient to work in the adimensional scaling variables
$\kappa^2={{\bf k}^2_\perp\over \hat{q}L}$, $\omega_c=
\hat{q}L^2/2$, $R=\omega_cL$, $\gamma = \omega_c/\omega$ and
${M}^2 = \frac{x^2 m^2}{\hat{q}L}$. In Fig.~\ref{fig2} the ${\bf
k}_\perp$-differential spectrum of radiated gluons is shown for different
gluon energies. For comparison, the massless result and the product of this
massless result times the dead cone factor $F\left({\bf k}^2_\perp,{m\omega
\over E}\right)$ are also shown. It can be seen
that the dead cone is filled, but also that it corresponds to a small fraction
of the available phase space. On the other hand, at large $\kappa$ the
radiation in the massive case is suppressed. Let us indicate that
only the sum of vacuum and medium pieces has to be positive.
In the massless case the vacuum contribution for $\kappa \to 0$
is positive and divergent, so
the medium contribution may become negative~\cite{sw}, while for
the massive case the dead cone effect kills the vacuum radiation for $\kappa
\to
0$ so the medium contribution cannot be negative.

\begin{figure}
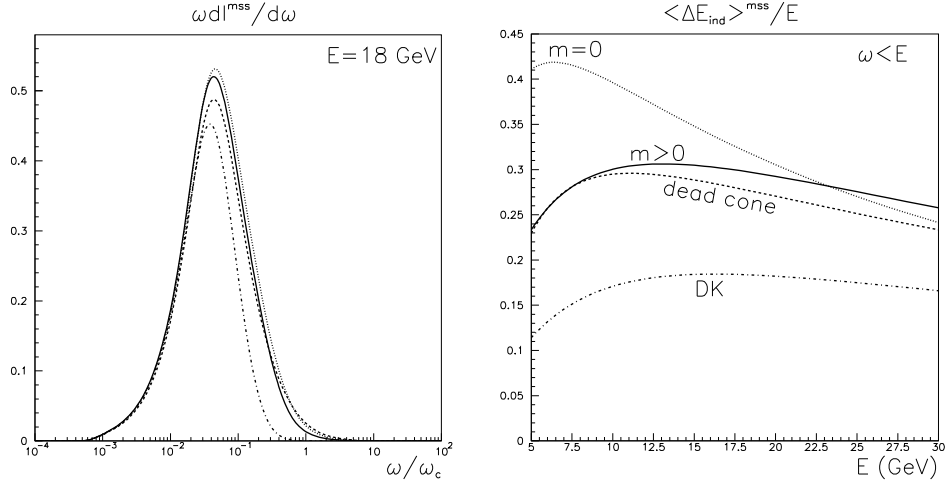

\begin{center}
\psfig{figure=fig9left.epsi,height=2.5in}
\hskip 0.5cm
\psfig{figure=fig9right.epsi,height=2.5in}
\end{center}
\caption{Left: Gluon energy distribution versus $\omega$ for a charm
quark ($m=1.5$ GeV) with energy
$E=18$ GeV.
Right: Fractional energy loss for different quark energies, for $L=6$ fm and
$\hat{q}=
0.8$ GeV$^2$/fm.
Solid (dotted)
lines are the results of the full massive (massless) computation of
Eq.~\protect{\ref{eq2}}, dashed lines are the massless results times the dead
cone factor $F\left({\bf k}^2_\perp,{m\omega
\over E}\right)$, and dashed-dotted lines are the results of
Eq.~\protect{\ref{eq1}} with $\langle{\bf
k}_\perp^2\rangle=\sqrt{\hat q \omega}$.
\label{fig4}}
\end{figure}

The ${\bf
k}_\perp$-integrated spectrum and the mean energy loss
are obtained as
\begin{equation}
\label{eq3}
\omega {dI_{\rm medium}\over d\omega}
=\int_0^\omega d{\bf k}_\perp\,
\omega\frac{dI_{\rm medium}}{d\omega\, d{\bf k}_\perp}\ ,
\ \ \langle \Delta E_{\rm ind}\rangle
  = \int_0^E d\omega\,   \omega\frac{dI_{\rm medium}}{d\omega}\ .
\end{equation}
It can be seen in Fig.~\ref{fig4} left that the ${\bf k}_\perp$-integrated
radiation is suppressed (in qualitative agreement with the dead cone
proposal~\cite{dk}). On the other hand, the result of Eq.~\ref{eq1} with
$\langle{\bf
k}_\perp^2\rangle=\sqrt{\hat q \omega}$ underestimates the emission.
Finally, in Fig.~\ref{fig4} right the mean energy loss
is shown for parameters taken from~\cite{sw}. For RHIC, $E\simeq 5\div 10$
GeV, the energy loss for charmed quarks is a factor $\sim 2$ smaller than that
for massless quarks, but should still be observable.
At higher energies, the energy loss in both cases tends
to the same value. Nevertheless, it can be observed a crossover between the
massive and massless cases. While it can be understood from
Fig.~\ref{fig4} left considering the moving upper integration limit in
Eq.~\ref{eq3}, it points out the uncertainties which are present in all
computations. Eq.~\ref{eq2} has been derived taking into account only leading
terms in $1/E$ ($x\ll 1$), so the kinematical limits implemented in
Eq.~\ref{eq3} are imposed {\it a posteriori} and lead to the
feature mentioned previously. As a last comment,
the results computed with the dead cone
factor agree quite closely with those of the full computation, while
Eq.~\ref{eq1} underestimates the energy loss.

\section{Conclusions}
We have computed the medium-induced
gluon radiation off massless and massive quarks
in the same formalism. Ours in the first ${\bf
k}_\perp$-differential result, consistent with available ${\bf
k}_\perp$-integrated ones~\cite{old}.
We find
that medium-induced gluon radiation fills the dead cone.
However,
the dead cone (i.e. the low-${\bf k}_\perp$
region) does not dominate the energy loss.
Our study suggests that energy loss for charmed hadrons at
RHIC should be
smaller than that for lighter hadrons, but still sizable (for
$p_\perp^{\rm hadron} \simeq 5\div 10$ GeV/c where hadronization effects
inside the medium should be negligible).
Finally,
the commented uncertainties motivate the computation of
$1/E$
corrections. In this way, the study of energy loss of massive quarks
(and of more differential observables~\cite{sw2})
offers new possibilities to check the existing formalism
and to restrict model parameters.

To conclude let us comment on the experimental situation. As of today, the
only experimental information about open charm
production in Au+Au collisions at RHIC is the prompt electron
spectrum measured by PHENIX~\cite{phenix}, which do not indicate a
significant parton energy loss for charmed hadrons but
also do not
constrain parton energy loss significantly (due to experimental errors,
a weak correlation between the transverse
momentum of the electron~\cite{muller}
and of the charmed hadron, and the low values of
$p_\perp^{\rm charm}$ which may be affected by hadronization).
The reconstruction of hadronic
decays of charmed hadrons will offer new possibilities (see~\cite{andrea}).

\section*{Acknowledgments}
We thank R. Baier, A. Dainese, K. Eskola,
H. Honkanen, A. Morsch, G. Rodrigo and J. Schukraft for helpful
discussions. N.A. also thanks
the organizers for their invitation to such a nice meeting.

\end{document}